\documentclass[%
 aip,
 groupedaddress,
 amsmath,amssymb,
 reprint,%
]{revtex4-1}

\usepackage{graphicx}% Include figure files
\usepackage{dcolumn}% Align table columns on decimal point
\usepackage{bm}% bold math
\usepackage{dsfont}
\usepackage{graphicx} 
\usepackage{xcolor}
\usepackage{comment}
\usepackage{lineno}
\usepackage{amsmath}

\makeatletter
\def\@email#1#2{%
 \endgroup
 \patchcmd{\titleblock@produce}
  {\frontmatter@RRAPformat}
  {\frontmatter@RRAPformat{\produce@RRAP{*#1\href{mailto:#2}{#2}}}\frontmatter@RRAPformat}
  {}{}
}%
\makeatother
\begin{document}

\preprint{AIP/123-QED}

\title[]{CUT-E as a $1/N$ expansion for multiscale molecular polariton dynamics}
% Force line breaks with \\
\author{Juan B. P\'erez-S\'anchez}
\author{Arghadip Koner}
\author{Sricharan Raghavan-Chitra}
\author{Joel Yuen-Zhou$^{*}$}%
 \email{joelyuen@ucsd.edu}
\affiliation{ 
Department of Chemistry, University of California San Diego, La Jolla, CA 92093, USA
}%

\date{\today}% It is always \today, today,

\begin{abstract}

Molecular polaritons arise when the collective coupling between an ensemble of $N$ molecules and an optical mode exceeds individual photon and molecular linewidths. The complexity of their description stems from their multiscale nature, where the local dynamics on each molecule can, in principle, be influenced by the collective behavior of the entire ensemble. To address this, we previously introduced a formalism called collective dynamics using truncated equations (CUT-E). CUT-E approaches the problem in two stages. First, it exploits permutational symmetries to obtain a substantial simplification of the problem. However, this is often insufficient for parameter regimes relevant to most experiments. Second, it takes the exact solution of the problem in the $N\to\infty$ limit as a reference and derives systematic $\mathcal{O}(1/N)$ corrections. Here we provide a novel derivation of CUT-E based on recently developed bosonization techniques. We lay down its connections with $1/N$ expansions that are ubiquitous in other fields of physics, and present previously unexplored key aspects of this formalism, including various types of approximations and extensions to high excitation manifolds.
\end{abstract}

\maketitle

\section{\label{sec:intro}Introduction}

Strong light-matter coupling is achieved when the coherent energy exchange rate between the degrees of freedom (DOF) of matter and a confined electromagnetic field surpasses the losses from either component, forming hybrid light-matter states known as polaritons \cite{ribero2018polariton,Ebbesen2023introduction,Wei2023what,Bhuyan2023rise,Mandal2023theoretical,Xiang2024molecular}. In molecular systems, strong coupling is typically attained through the interaction of an ensemble of $N\gg 1$ molecules with a microcavity mode due to the small magnitude of the transition dipole moment of individual molecules. Over the past decade, microcavities have garnered significant attention for their potential for enhanced energy and charge transport \cite{coles2014polariton,Reitz2018energy,delpo2021polariton,martinez2018polariton,xiang2020intermolecular,koner2023path,Rabani2024polariton}, modification and control of chemical reactions \cite{hutchison2012modifying,dunkelberger2016modified,zeng2023control,Dunkelberger2022vibration,Weichman2024more}, and room-temperature polariton condensation \cite{Ishii2022low,kena2010room}, among other remarkable phenomena. However, many of these effects remain unresolved, largely due to the computational challenges posed by the multiscale nature of the systems in question. This theoretical complication arises as the collective behavior of the ensemble can influence the local dynamics of each molecule. As a result, most existing molecular polariton simulations are limited to a few tens of molecules using sophisticated numerical methods \cite{groenhof2017multiscale, groenhof2019tracking,Tao2022energy,Tao2023QMMM} or single- or few-molecule systems using \textit{ab initio} techniques \cite{rubio2021polaritonic,rubio2022perspective,rubio2022polaritonic,Koch2024toward}.

To address these challenges, we recently introduced a formalism called collective dynamics using truncated equations (CUT-E) \cite{Perez2022simulating}. First, it exploits permutational symmetry for an efficient representation of the system. Second, it provides a simple solution to the problem for large $N$ by deriving an exact solution in the $N\to\infty$ limit and subsequently carrying out systematic $1/N$ corrections. This resembles many other techniques in different areas of physics. \cite{witten1979instantons,Hooft1974largenqcd,chatterjee1990largeN,Witten1980quarks,Vollhardt2012dmft}. One of the most seminal instances is the 't Hooft's large-$N$ expansion that simplifies the analysis of color group $U(N)$ gauge theories, and established a duality between quantum chromodynamics (QCD) and string theory in the large-$N$ limit~\cite{Hooft1974largenqcd}. The use of this top-down approach to study molecular polaritons has recently grown in popularity \cite{zeb2018exact,Piper2022PRL,Cui2022JCP,Keeling2022incoherent,Gu,Sidler2024connection,Lindoy2024investigating}, as it brings substantial computational efficiency compared to the traditional bottom-up methodologies often applied to mesoscopic ensembles.
Our original derivation of CUT-E was carried out in a pedestrian fashion using a first quantization approach to the many-molecule plus cavity wavefunction. Here, we provide a much simpler re-derivation using a recently developed bosonic mapping of polaritons and make new direct connections of CUT-E with $1/N$ expansions of other fields.

The article is organized as follows: in Section \ref{sec:model}, we present the Tavis-Cummings-Holstein Hamiltonian and establish a bosonic mapping to the vibronic modes of the single molecule. In section \ref{sec:colletive}, we state the difference between the collective \textit{vs} single-molecule light-matter coupling regimes. In section \ref{sec:quantum}, we rigorously solve the quantum dynamics in the first excitation manifold for the case of $N\to\infty$. Then, we show that exact quantities can be calculated for finite $N$ via a $1/N$ expansion. In section \ref{sec:1nexp}, we present our method in a broader setting and connect it to $1/N$ expansions in other areas of physics. In section VI, we briefly comment on the open problems in the polaritonic quantum dynamics in high excitation manifolds. Sections \ref{sec:model}, \ref{sec:colletive}, and \ref{sec:quantum} yield new insights and simpler re-derivations of results that have been already reported \cite{Perez2022simulating,Perez2023collective,Koner2024linear,Perez2024radiative,Kai2024filtering}, while sections \ref{sec:1nexp} and \ref{sec:high} contain entirely new results.

\section{\label{sec:model}Model}

\subsection{\label{sec:wf}First-quantized picture}

The Hamiltonian for an ensemble of non-interacting molecules strongly coupled to a confined optical mode of a microcavity can be written as
\begin{equation}\label{eq:ham}
\hat{H}=\sum_{i}^{N}\left(\hat{H}_{mol}^{(i)}+\hat{H}^{(i)}_{I}\right)+\hat{H}_{cav},
\end{equation} where $\hat{H}_{mol}^{(i)}$ is the molecular Hamiltonian of the $i$th molecule, $\hat{H}_{cav}$ is the Hamiltonian for the cavity modes, and $\hat{H}^{(i)}_{I}$ is the light-matter interaction between the $i$th molecule and the cavity modes.

For simplicity, let us consider all molecules to be identical and the cavity to have a single photon mode. Although these assumptions are important for our method to work, we discuss how to generalize them to account for disorder and the multimode nature of the cavity further below. Also, for pedagogical purposes, let us assume the Born-Oppenheimer approximation for the molecular Hamiltonian and the Condon and rotating wave approximations to the light-matter interaction (these approximations can be easily relaxed). Thus, the molecular polariton Hamiltonian can be written as
\begin{align}\label{eq:hamm}
\hat{H}&=\omega_{c}\hat{a}^{\dagger}\hat{a}+\sum_{i}^{N}\left(\hat{T}_{i}+V_{g}(q_{i})|g_{i}\rangle\langle g_{i}|+V_{e}(q_{i})|e_{i}\rangle\langle e_{i}|\right)\nonumber\\
&+g\sum_{i}^{N}\left(|e_{i}\rangle\langle g_{i}|\hat{a}+|g_{i}\rangle\langle e_{i}|\hat{a}^{\dagger}\right),
\end{align} where $\hat{T}$ is the kinetic energy operator, $V_{g/e}$ are the ground/excited potential energy surfaces (PES), $g$ is the single-molecule light-matter coupling strength, and $\hat{a}$ is the annihilation operator of a photon in the cavity mode. 
%=-\sqrt{\frac{\omega_{c}}{2\epsilon_{0}\mathcal{V}}}\mu_{eg}

Notice that $\hat{H}$ is invariant under the permutation of any pair of molecules. For a permutationally-symmetric initial wavefunction, this symmetry is conserved throughout the time evolution. This means we can write the \textit{exact} time-dependent many-body wavefunction of the polariton system as a multiconfigurational expansion 
\begin{widetext}
\begin{equation}\label{eq:fqmbwf}
\Psi(q_{ph},\vec{q},t)=\sum_{n_{ph}}\sum_{\nu_{1}\nu_{2}\cdots \nu_{N}}\tilde{A}^{(n_{ph})}_{\nu_{1}\nu_{2}\cdots \nu_{N}}(t)\varphi_{n_{ph}}(q_{ph})\hat{S}_{+}\prod_{k=1}^{N}\varphi_{\nu_{k}}(q_{k}),
\end{equation}
\end{widetext} where $\varphi_{n_{ph}}$ are the Fock states for the photon mode, $\varphi_{\nu_{k}}(q_{k})$ is the $\nu_{k}$th single-particle basis state of the $k$th molecule,  and $\hat{S}_{+}$ is the bosonic symmetrization operator. In other words, the dynamics of the system can be computed in a permutationally-symmetric subspace of the entire Hilbert space.

\subsection{\label{sec:bosonic}Second-quantized picture}

Our original paper on the CUT-E method is based on the first-quantized many-body wavefunction in Eq. \ref{eq:fqmbwf}. However, it involves cumbersome symmetrization and renormalization procedures that can be bypassed by working in a more natural second-quantized picture from the onset \cite{Richter2016efficient,Shammah2018open,KeelingZeb2,Silva2022permutational,Pizzi2023light,Sukharnikov2023second}. Let $\hat{O}=\sum_{k=1}^{N}\hat{o}^{(k)}$ be the total (symmetric) sum of single-molecule operators $\hat{o}^{(k)}$, we can move to a second-quantized picture via the mapping
\begin{equation}
    \sum_{k=1}^{N}\hat{o}^{(k)}\rightarrow \sum_{\nu_{i}\nu_{j}}\langle\nu_{i}|o|\nu_{j}\rangle \hat{\beta}^{\dagger}_{\nu_{i}}\hat{\beta}_{\nu_{j}},
\end{equation} where $\hat{\beta}_{i}$ is a bosonic operator that annihilates a molecule in the \textit{single-molecule} vibronic state $|\nu_{i}\rangle$ (see Appendix A).

By applying the above mapping to the Hamiltonian in Eq. \ref{eq:hamm}, we obtain the bosonic molecular polariton Hamiltonian
\begin{align}\label{eq:ham2}
\hat{H}&=\omega_{c}\hat{a}^{\dagger}\hat{a}+\sum_{i}^{m}\omega_{g,i}\hat{b}_{i}^{\dagger}\hat{b}_{i}+\sum_{i}^{m}\omega_{e,i}\hat{B}_{i}^{\dagger}\hat{B}_{i}\nonumber\\
&+g\sum_{ij}^{m}\left(\langle\varphi^{(e)}_{i}|\varphi^{(g)}_{j}\rangle\hat{B}^{\dagger}_{i}\hat{b}_{j}\hat{a}+\langle\varphi^{(g)}_{j}|\varphi^{(e)}_{i}\rangle\hat{B}_{i}\hat{b}^{\dagger}_{j}\hat{a}^{\dagger}\right).
\end{align} Here, $\hat{b}_{i}$ and $\hat{B}_{j}$ annihilate a \textit{molecule} in the vibronic basis states $|\varphi^{(g)}_{i},g\rangle$ and $|\varphi^{(e)}_{j},e\rangle$. Due to the BO approximation, there are no couplings between $\hat{b}_{i}$ and $\hat{B}_{j}$ (they are eigenstates). Moreover, $m$ is the number of vibrational basis states, $\omega_{g/e,i}$ are the vibronic eigenvalues, and $\langle\varphi^{(e)}_{i}|\varphi^{(g)}_{j}\rangle$ are the Franck-Condon factors of the corresponding optical transition. This mapping can be regarded as a generalization of the Schwinger boson representation for spins to systems with more than two states \cite{schwinger1965boson}, and has been previously employed in the context of molecular polaritons by several authors \cite{Richter2016efficient,Shammah2018open,Keeling2022efficient,Silva2022permutational,Pizzi2023light,Sukharnikov2023second,Perez2024radiative}. Yet, we must keep in mind that it only describes the dynamics of the system for permutationally-symmetric Hamiltonians and initial conditions.

\begin{figure}[htb]
    \centering
    \includegraphics[width=1\linewidth]{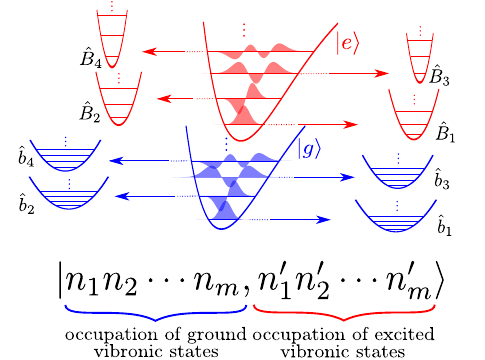}
    \caption{Boson mapping of the molecular polariton Hamiltonian. Molecules are effectively treated as bosonic particles that occupy single-molecule vibronic states. This mapping is exact for an ensemble of identical molecules, and for permutationally symmetric initial conditions. Here we use the molecular eigenstates $|\varphi^{(g)}_{i},g\rangle$ and $|\varphi^{(e)}_{i},e\rangle$ as the basis set (see Eq. \ref{eq:ham2}).}
    \label{fig:fig2}
\end{figure}

The many-body basis states $|n_{1}n_{2}\cdots n_{m},n^{\prime}_{1}n^{\prime}_{2}\cdots n^{\prime}_{m},n_{ph}\rangle$ are eigenstates of the non-interacting Hamiltonian (i.e., for $g=0$), with $n_{ph}$ being the number of photons in the cavity, and $n_{i}$ and $n^{\prime}_{i}$ being the number of molecules in the $|\varphi^{(g)}_{i},g\rangle$ and $|\varphi^{(e)}_{i},e\rangle$ vibronic states, respectively (see Fig. \ref{fig:fig2}). These states do not track information of the dynamics of each molecule; instead, they describe how many molecules are in each available state. This contrasts brute-force molecular dynamics simulations that track the vibronic state of each of the $N$ molecules, leading to computational costs that grow exponentially with $N$.

It can be easily checked that the molecular polariton Hamiltonian conserves the number of molecules and the number of excitations by noticing that $\hat{N}_{mol}=\sum_{i}^{m}(\hat{b}^{\dagger}_{i}\hat{b}_{i}+\hat{B}^{\dagger}_{i}\hat{B}_{i})$, $\hat{N}_{exc}=\sum_{i}^{m}\hat{B}^{\dagger}_{i}\hat{B}_{i}+\hat{a}^{\dagger}\hat{a}$, and $\left[\hat{N}_{mol},\hat{H}\right]=\left[\hat{N}_{exc},\hat{H}\right]=0$. Moreover, generalization to include nonadiabatic couplings, non-Condon effects, and counter-rotating light-matter coupling terms is simple as such terms do not break the permutational symmetry of the Hamiltonian in Eq. \ref{eq:ham}.

Notice that this bosonic representation alone is not enough to allow for the exact simulation of the system's dynamics. In fact, quantum dynamics simulations in the bosonic picture continue to be intractable for the number of molecules, excitations, and vibronic states involved in many experiments. This will be addressed in the next sections. 

\section{\label{sec:colletive}Collective vs single-molecule light-matter coupling}

\begin{figure}[htb]
    \centering
    \includegraphics[width=1\linewidth]{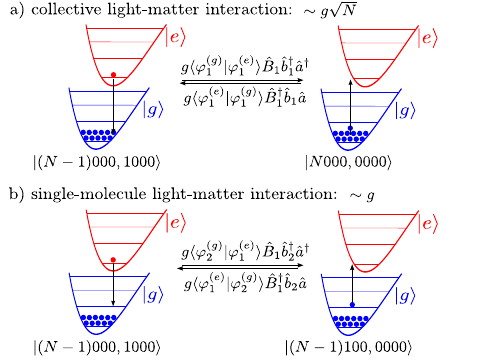}
    \caption{Collective vs. single-molecule light-matter coupling terms. At zero temperature, collective coupling involves absorption and emission processes that do not create phonons (e.g., Rayleigh scattering), while single-molecule coupling involves processes that do (e.g., fluorescence). See Eq. \ref{eq:ham2}.}
    \label{fig:fig3}
\end{figure}

Consider an initial state with $N-1$ molecules in the global ground state and a single excited molecule (zero temperature), i.e., $|\Psi(0)\rangle=|(N-1)0\cdots 0,10\cdots 0,0\rangle$. We can obtain intuition about the dynamics of the excited molecule in the cavity by analyzing the light-matter coupling terms between $|\Psi(0)\rangle$ and the other many-body basis states. Notably, light-matter couplings that destroy the excited state molecule while creating a cavity photon and a molecule at the macroscopically occupied \textit{global ground state} are amplified by a factor of $\sqrt{N}$, while light-matter couplings that create molecules in the unoccupied vibrationally excited states of the ground electronic state are not (see Fig. \ref{fig:fig3}). This collective light-matter interaction can be understood as a consequence of bosonic stimulation, although it can also be thought of as a consequence of the constructive interference of emission pathways for all the molecules when starting with a permutationally-symmetric excited state \cite{Perez2022simulating}.

The clear disparity between collective and single-molecule coupling strengths in the large $N$ limit motivates a partitioning of the bosonic molecular polariton Hamiltonian as $\hat{H}=\hat{H}^{(0)}+\hat{v}$, with
\begin{align}\label{eq:part}
\hat{H}^{(0)}&=\omega_{c}\hat{a}^{\dagger}\hat{a}+\sum_{i}^{m}\omega_{g,i}\hat{b}_{i}^{\dagger}\hat{b}_{i}+\sum_{i}^{m}\omega_{e,i}\hat{B}_{i}^{\dagger}\hat{B}_{i}\nonumber\\
&+g\sum_{i}^{m}\left(\langle\varphi^{(e)}_{i}|\varphi^{(g)}_{1}\rangle\hat{B}^{\dagger}_{i}\hat{b}_{1}\hat{a}+\langle\varphi^{(g)}_{1}|\varphi^{(e)}_{i}\rangle\hat{B}_{i}\hat{b}^{\dagger}_{1}\hat{a}^{\dagger}\right)\nonumber\\
&\hat{v}=g\sum_{i=1,j>1}^{m}\left(\langle\varphi^{(e)}_{i}|\varphi^{(g)}_{j}\rangle\hat{B}^{\dagger}_{i}\hat{b}_{j}\hat{a}+\langle\varphi^{(g)}_{j}|\varphi^{(e)}_{i}\rangle\hat{B}_{i}\hat{b}^{\dagger}_{j}\hat{a}^{\dagger}\right).
\end{align} This partitioning will be instrumental in the next sections.

Naturally, a large number of cavity photons can enhance the light-matter interaction as well (e.g., via stimulated emission). So care must be taken when utilizing this partitioning when a large number of excitations are present.

\section{\label{sec:quantum}Quantum dynamics in the first excitation manifold}

Here we focus on the Hilbert subspace with $N_{exc}=1$. This so-called first excitation manifold describes the system containing a single excitation either as a photon or as an electronic excitation. This regime is very important as it describes approximately the polariton system being irradiated by a weak laser \cite{Sukharev2023comparing} (the linear regime in the external laser field). 

Our method, called collective dynamics using truncated equations (CUT-E) \cite{Perez2022simulating}, was developed to study the first-excitation manifold. At the heart of the method is the observation that the collective couplings are much larger than the single-molecule counterparts ($N\gg 1$). CUT-E reveals a structure of the equations of motion of the molecular polariton system that allows for their systematic truncation. Our original derivation was carried out in the language of first quantization; here, we provide a new and simpler derivation using the bosonic representation discussed in the previous sections. For simplicity, let us use the multi-particle states notation introduced by Philpott \cite{Philpott} in the context of electronic spectra of molecular aggregates, which has been used to describe molecular polaritons previously by Herrera and Spano \cite{Herrera2017Dark},

\begin{align}
    &|1\rangle=|N00\cdots 0,00\cdots 0,1\rangle\nonumber\\
    &|e_{k}\rangle=|(N-1)00\cdots 0,\cdots 1_{k}\cdots,0\rangle\nonumber\\
    &|g_{k}1\rangle=|(N-1)\cdots 1_{k}\cdots,00\cdots 0,1\rangle\nonumber\\
    &|g_{k}e_{k'}\rangle=|(N-2)\cdots 1_{k}\cdots,\cdots 1_{k'}\cdots,0\rangle\nonumber\\
    &|g_{k}g_{k'}1\rangle=|(N-2)\cdots 1_{k}\cdots 1_{k'}\cdots,00\cdots 0,1\rangle\nonumber\\
    &|g_{k}g_{k'}e_{k''}\rangle=|(N-2)\cdots 1_{k}\cdots 1_{k'}\cdots,\cdots 1_{k''}\cdots 0,0\rangle\nonumber\\
    &\hspace{3.5cm}\vdots
\end{align}

Next, we notice that the molecular polariton Hamiltonian of Eq. \ref{eq:ham2} becomes block-tridiagonal in this basis (see Appendix B), 
\begin{equation}\label{eq:blocks}
\boldsymbol{H}=\begin{pmatrix}
\boldsymbol{H}^{(0)}_{0} & \boldsymbol{v}_{0} & 0 & \cdots & 0 \\
\boldsymbol{v}^{\dagger}_{0} & \boldsymbol{H}^{(0)}_{1} & \boldsymbol{v}_{1} & \cdots & 0 \\
0 & \boldsymbol{v}^{\dagger}_{1} & \boldsymbol{H}^{(0)}_{2} & \cdots & 0 \\
\vdots & \vdots & \vdots & \ddots & \vdots \\
0 & 0 & 0 & \cdots & \boldsymbol{H}^{(0)}_{N} \\
\end{pmatrix}. 
\end{equation} Here, we have used the partitioning of the Hamiltonian in Eq. \ref{eq:part}, where all collective light-matter couplings are contained inside each block $\boldsymbol{H}^{(0)}_{i}$, while the single-molecule light-matter interaction $\boldsymbol{v}$ couple neighboring blocks. The subscript $i$ corresponds to the number of electronic ground state molecules with vibrational excitations, and we refer to it as a \textit{quasi}-conserved quantity. In other words, only vibrational excitations in electronic ground state molecules can be created only via the (small) single-molecule light-matter coupling $\boldsymbol{v}$. The states $|1\rangle$ and $|e_{k}\rangle$ form a complete basis set for $\hat{H}^{(0)}_{0}$, the states $|g_{k}1\rangle$ and $|g_{k}e_{k'}\rangle$ form a complete basis set for $\hat{H}^{(0)}_{1}$, and so on. 

\begin{figure}[htb]
    \centering
    \includegraphics[width=1\linewidth]{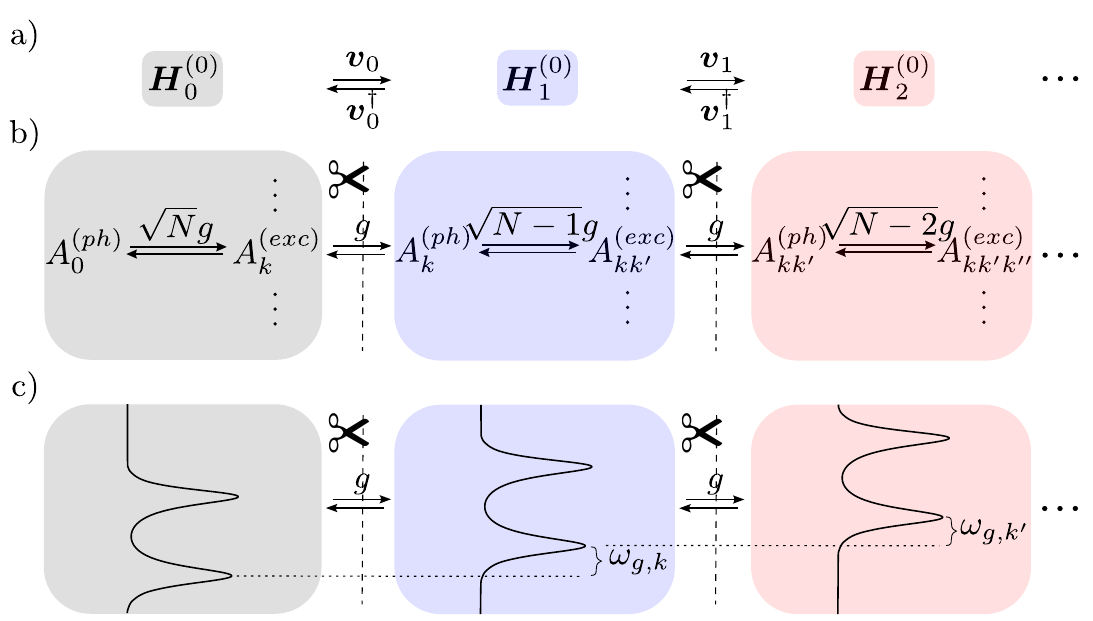}
    \caption{Structure of the molecular polariton Hamiltonian. a) Block-tridiagonal structure of the Hamiltonian in Eq. \ref{eq:blocks}. b) Each block contains all vibronic couplings and collective light-matter couplings while coupling between blocks is single-molecule-like. c) Within each block, we can find vibronic polariton and dark states. For $N\gg 1$, the states in one block are simply shifted by the energy of a vibrational excitation with respect to those of the previous block (see Sec. \ref{sec:largen}).}
    \label{fig:fig4}
\end{figure}

The CUT-E method approximates the polariton dynamics by truncating the basis of the ansatz wavefunction
\begin{align}\label{eq:wf}
    |\Psi(t)\rangle&=A^{(ph)}_{0}(t)|1\rangle+\sum_{k=1}^{m}A^{(exc)}_{k}(t)|e_{k}\rangle+\sum_{k>1}^{m}A^{(ph)}_{k}(t)|g_{k}1\rangle\nonumber\\
    &+\sum_{k>1,k^{\prime}=1}^{m}A^{(exc)}_{kk'}(t)|g_{k}e_{k'}\rangle+\sum_{k,k'>1}^{m}A^{(ph)}_{kk'}(t)|g_{k}g_{k'}1\rangle\nonumber\\
    &+\sum_{k,k'>1,k''=1}^{m}A^{(exc)}_{kk'k''}(t)|g_{k}g_{k'}e_{k''}\rangle+\cdots
\end{align} For instance, if only collective light-matter coupling terms are to be accounted for, only the first two terms in the RHS of Eq. \ref{eq:wf} must be included in the ansatz wavefunction. We call this zeroth-order CUT-E. On the other hand, if \textit{at least} one action of the single-molecule coupling is to be considered, the first four terms must be included. We call this first-order CUT-E. This hierarchy is summarized in the scheme of Fig. \ref{fig:fig4}.

\subsection{\label{sec:ninf}The $N\rightarrow\infty$ limit}

A simple solution to the molecular polariton dynamics is obtained by taking the limit when $N\rightarrow\infty$ (or $g\rightarrow 0$) while keeping the collective coupling $g\sqrt{N}$ constant. In this case, the Hamiltonian in Eq. \ref{eq:blocks} becomes block diagonal, and the problem is trivially solved by diagonalizing each block separately. The eigenstates of these blocks can be used as reference states from which the exact eigenstates for finite $N$ can be approximated. As an illustrative example, let us analyze the zeroth block (hereafter $\omega_{g,1}=0$),
\begin{widetext}
\begin{equation}\label{eq:h0}
\boldsymbol{H}^{(0)}_{0}=\begin{pmatrix}
\omega_{c} & g\sqrt{N}\langle\varphi^{(g)}_{1}|\varphi^{(e)}_{1}\rangle & g\sqrt{N}\langle\varphi^{(g)}_{1}|\varphi^{(e)}_{2}\rangle & \cdots & g\sqrt{N}\langle\varphi^{(g)}_{1}|\varphi^{(e)}_{m}\rangle \\
g\sqrt{N}\langle\varphi^{(e)}_{1}|\varphi^{(g)}_{1}\rangle & \omega_{e,1} & 0 & \cdots & 0 \\
g\sqrt{N}\langle\varphi^{(e)}_{2}|\varphi^{(g)}_{1}\rangle & 0 & \omega_{e,2} & \cdots & 0 \\
\vdots & \vdots & \vdots & \ddots & \vdots \\
g\sqrt{N}\langle\varphi^{(e)}_{m}|\varphi^{(g)}_{1}\rangle & 0 & 0 & \cdots & \omega_{e,m}
\end{pmatrix}.
\end{equation}   
\end{widetext}

\begin{figure}[htb]
    \centering
    \includegraphics[width=1\linewidth]{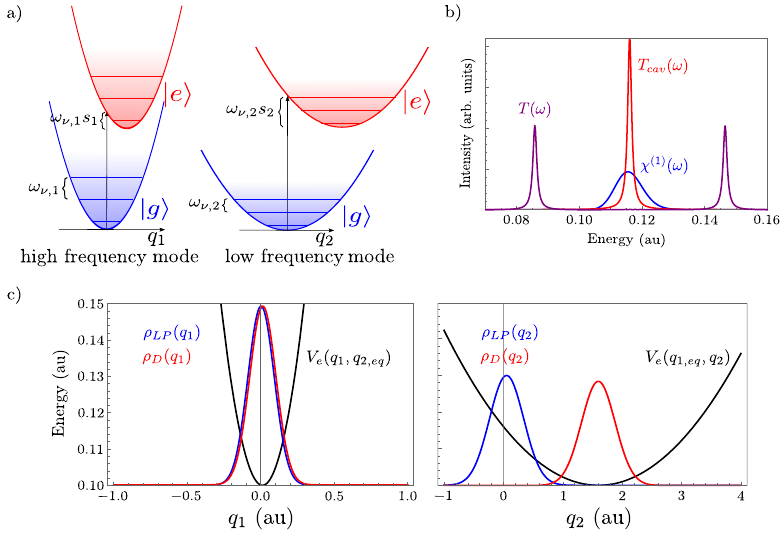}
    \caption{Polariton linear response showcasing vibronic polaritons and dark states. a) Linear vibronic coupling model with high- and low-frequency vibrational modes. b) Cavity, bare molecular, and polariton absorption spectra. c) Lower polariton and lowest energy dark state vibronic densities along each vibrational coordinate $\rho_{LP}(q_{1,2})$ and $\rho_{D}(q_{1,2})$ (the densities are re-scaled and plotted on top of the PESs for illustration purposes). While the lower polariton state corresponds to the excited molecules near the Franck-Condon region, the lowest energy dark state corresponds to an excited state vibrationally-relaxed (Stokes-shifted) configuration. In these calculations we used $\omega_{\nu,1}=10\omega_{\nu,2}=0.01$ au, $s_{1}=0.01$, $s_{2}=16$, $\omega_{c}=\omega_{0}+\omega_{\nu,1}s_{1}+\omega_{\nu,2}s_{2}=0.1161$ au, $g\sqrt{N}=0.03$ au, and cavity leakage $\kappa=0.0015$ au.}
    \label{fig:fig5}
\end{figure}

In this picture, the \textit{vibro-polaritonic} eigenstates of $\boldsymbol{H}^{(0)}_{0}$ with a high and negligible photonic character are called ``polariton'' and ``dark states'' states, respectively. Polaritons typically have a large contribution from the bright state $|\varphi^{(g)}_{1},e\rangle$ and form two bands known as the upper (high frequency) and lower (low frequency) polaritons. On the other hand, dark states are Stokes-shifted states that remain at their original energies outside the cavity (see Fig \ref{fig:fig5}b).

Besides the obvious simplification obtained by having to calculate the eigenstates of only a single block, using $\boldsymbol{H}^{(0)}_{0}$ to model the polaritonic system is equivalent to using classical linear optics treatments, where the photonic mode of frequency $\omega_{c}$ is strongly coupled to a set of effective oscillators representing the electronic transitions of the bare molecule with frequencies $\omega_{e,i}$, which can be obtained from the linear response of the bare molecular ensemble \cite{Cwik2016excitonic,Keeling2018exact,Koner2024linear,Manolopoulos2023vibrational}. In fact, it was recently pointed out that the molecular dynamics under $\boldsymbol{H}^{(0)}_{0}$ can be reproduced by a weak laser $E(t)\propto D_{N\rightarrow \infty}^{(R)}(t)=-i\Theta(t)\langle 1|e^{-i\hat{H}^{(0)}_{0}t}|1\rangle$ acting on the bare molecular ensemble \cite{Kai2024filtering}. In the frequency domain, the polaritons act mainly as optical filters that allow only certain frequencies to enter the cavity and interact weakly with the molecular ensemble \cite{Kai2024filtering}.

This theoretical prediction holds even in the presence of static disorder, where permutational symmetries are broken by small variations in the molecular environments. For CUT-E to be implemented in this scenario, we coarse-grain the disorder distribution and apply permutational symmetries to the molecules within each disorder bin. Since the ability of the cavity to resolve energetic differences between molecules depends on the propagation time, a relatively small number of disorder bins $N_{bins}$ is needed to capture the ultrafast dynamics of the disorder ensemble, and we can take the limit where the number of molecules within each disorder bin goes to infinity \cite{Perez2024collective}. Both the numerical convergence of the dynamics using this coarse-graining approach \cite{Sokolovskii2024one,Perez2024collective}, and the optical filtering prediction \cite{Dutta2024thermal} have been numerically tested in recent works. We believe a similar approach can be used to incorporate multiple cavity modes in the CUT-E formalism by coarse-graining the distribution of molecules in real space, which can open a path towards the efficient simulation of polariton transport and polariton condensation in the large $N$ limit. Recent theoretical works have already made significant progress in this direction \cite{Cao2023polariton}.

\subsection{\label{sec:largen}$1/N$ corrections}

From the previous discussion, it becomes clear that, within the limitations of this theory (e.g., no distance-dependent intermolecular interactions, finite temperature, etc), single-molecule light-matter coupling effects must be included in order to obtain polaritonic phenomena beyond those described by classical linear optics. Here, we show that these effects can be incorporated in a relatively simple manner using perturbation theory. 

As an example, consider an initial state $|i\rangle$ that lives in the zeroth block $\hat{H}^{(0)}_{0}$ (e.g., a dark state or a polariton at zero temperature). Using the partitioning of the Hamiltonian in Eq. \ref{eq:part}, we can express the \textit{survival amplitude} 
$c(t)=\langle i|e^{-i\hat{H} t} |i\rangle$ as a $1/N$ expansion:
\begin{widetext}
\begin{align}\label{eq:1n}
    c(t)&=\langle i|e^{-i\hat{H}^{(0)}_{0} t} |i\rangle-\int_{0}^{t}dt_{1}\int_{0}^{t_{1}}dt_{2}\langle i|e^{i\hat{H}^{(0)}_{0} t_{1}}\hat{v}^{\dagger}_{0}e^{-i\hat{H}^{(0)}_{1} (t_{1}-t_{2})}\hat{v}_{0}e^{-i\hat{H}^{(0)}_{0} t_{2}}|i\rangle+\cdots\nonumber\\
    c(t)&=c_{1}(t)+c_{1/N}(t)+\cdots= c_{1}(t)+\tilde{c}_{1/N}(t)\left(\frac{1}{N}\right)+\cdots.
\end{align}
\end{widetext} Notice that, due to the choice of the initial state and the block-tridiagonal structure of the Hamiltonian, only even actions of $\hat{v}$ are allowed. Moreover, each action of $\hat{v}$ adds a factor of $\sim 1/\sqrt{N}$, meaning that the perturbative expansion leads to a $1/N$ expansion, with $c_{1}(t),\tilde{c}_{1/N}(t),\dots$ in Eq. \ref{eq:1n} being $\mathcal{O}(1)$.

Although numerical calculations of such corrections can be challenging, a simplification is allowed in the limit where $N\gg 1$. First, notice that the second block from Eq. \ref{eq:blocks} can be written as $\boldsymbol{H}^{(0)}_{1}=\bigoplus_{k>1}^{m}\boldsymbol{H}^{(0)}_{1_{k}}$, with 
\begin{widetext}
\begin{equation}\label{eq:h1}
\boldsymbol{H}^{(0)}_{1_{k}}=\begin{pmatrix}
\omega_{c}+\omega_{g,k} & g\sqrt{N-1}\langle\varphi^{(g)}_{1}|\varphi^{(e)}_{1}\rangle & g\sqrt{N-1}\langle\varphi^{(g)}_{1}|\varphi^{(e)}_{2}\rangle & \cdots & g\sqrt{N-1}\langle\varphi^{(g)}_{1}|\varphi^{(e)}_{m}\rangle \\
g\sqrt{N-1}\langle\varphi^{(e)}_{1}|\varphi^{(g)}_{1}\rangle & \omega_{e,1}+\omega_{g,k} & 0 & \cdots & 0 \\
g\sqrt{N-1}\langle\varphi^{(e)}_{2}|\varphi^{(g)}_{1}\rangle & 0 & \omega_{e,2}+\omega_{g,k} & \cdots & 0 \\
\vdots & \vdots & \vdots & \ddots & \vdots \\
g\sqrt{N-1}\langle\varphi^{(e)}_{m}|\varphi^{(g)}_{1}\rangle & 0 & 0 & \cdots & \omega_{e,m}+\omega_{g,k}
\end{pmatrix}.
\end{equation}    
\end{widetext} 

Also, notice that the Hamiltonian in Eq. \ref{eq:h1} differs from that in Eq. \ref{eq:h0} in a constant energy shift by $\omega_{g,k}$ (the energy of the vibrational excitation in one ground state molecule), and in the reduced collective coupling $g\sqrt{N-1}$. In the $N\gg 1$ limit, we can approximate $\boldsymbol{H}^{(0)}_{1_{k}}\approx\boldsymbol{H}^{(0)}_{0}+\omega_{g,k}\boldsymbol{I}$. This implies that only the eigenstates and eigenvalues of $\boldsymbol{H}^{(0)}_{0}$ must be calculated to obtain the eigenstates of $\boldsymbol{H}^{(0)}_{1}$. A similar argument applies for all $\boldsymbol{H}^{(0)}_{i}$ blocks as long as low-order corrections are needed (see Fig. \ref{fig:fig4}c). Single-molecule coupling effects can be easily included via perturbation theory. The single-molecule light-matter coupling matrix $\boldsymbol{v}_{0}=\left(\boldsymbol{v}_{0,2}\ \boldsymbol{v}_{0,3}\ \cdots \ \boldsymbol{v}_{0,m}\right)$, where the matrix $\boldsymbol{v}_{0,k}$ that couples $\boldsymbol{H}^{(0)}_{0}$ with the \textit{sub-block} $\boldsymbol{H}^{(0)}_{1_{k}}$ is given by
\begin{equation}\label{eq:v0}
\boldsymbol{v}_{0,k}=\begin{pmatrix}
0 & 0 & 0 & \cdots & 0 \\
g\langle\varphi^{(e)}_{1}|\varphi^{(g)}_{k}\rangle & 0 & 0 & \cdots & 0 \\
g\langle\varphi^{(e)}_{2}|\varphi^{(g)}_{k}\rangle & 0 & 0 & \cdots & 0 \\
\vdots & \vdots & \vdots & \ddots & \vdots \\
g\langle\varphi^{(e)}_{m}|\varphi^{(g)}_{k}\rangle & 0 & 0 & \cdots & 0
\end{pmatrix}.
\end{equation} 

The perturbative approach in the $N\gg 1$ limit was recently used to describe the radiative decay of dark states \cite{Michetti2008simulation,coles2014polariton,Michetti2009exciton,coles2011vibrationally}. In first-order perturbation theory, we find the radiative pumping rate, $\mathcal{O}(1/N)$; while in second-order perturbation theory, we find the polariton-assisted Raman scattering rate, $\mathcal{O}(1/N^{2})$ \cite{Perez2024radiative}. The existence of a $1/N$ expansion implies that there are additional rate contributions beyond $1/N^{2}$ whose underlying physical mechanisms have not been explored yet.

\section{\label{sec:1nexp}CUT-E as a $1/N$ expansion}

Studying the properties of physical systems in the limit where one of its parameters $N$ is taken to infinity, followed by the addition of $1/N$ corrections, is a common practice in quantum field theories \cite{chatterjee1990largeN,witten1979instantons,Witten1980quarks} and the study of quantum-classical correspondence \cite{prosen2003echo,morita2022extracting,pappalardi2020sherrington,Hashimoto2017OTOC,Karnakov2013expansion}. For example, the limit where the number of field components approaches infinity \cite{Moshe2003quantum} usually leads to a classical description where quantum fluctuations are suppressed (as in Sec. \ref{sec:ninf}). Subsequently, $1/N$ corrections are systematically added to account for quantum fluctuations (as in Sec. \ref{sec:largen}). In electronic structure theory, classical electrostatics or mean-field theory become exact methods in the limit where the number of dimensions, electrons, and coordination number are taken to infinity \cite{Dunn1994a,Vollhardt2012dmft,Metzner1989correlated,Georges1996dmft}. In QCD, the number of colors is the parameter taken to infinity \cite{Hooft1974largenqcd,Manohar1998largenqcd}. The linear harmonic regime in the large $N$ limit is also found in quantum impurity models \cite{Makri1999linear}, and in the Sachdev-Ye-Kitaev (SYK) \cite{sachdev1993gapless,kitaev2015holography} and the Sherrington-Kirkpatrick \cite{Rosenhaus2019SYK,pappalardi2020sherrington} models in many-body physics where $1/N$ plays the role of the effective Planck’s constant. By mapping the polaritons problem to a quantum impurity model, it can be shown that the dynamics of the photon mode in the large $N$ limit can be exactly captured by replacing the complex anharmonic molecular bath with a surrogate harmonic one (see Eq. \ref{eq:h0}) \cite{Koner2024linear,chin2010impurity}. In our CUT-E method, each term of the $1/N$ expansion has its origins in single-molecule light-matter coupling processes. CUT-E teaches us that there is still much we can learn from $1/N$ expansions for otherwise intractable chemical systems.

\section{\label{sec:high}Quantum dynamics in high excitation manifolds}

Generalization of CUT-E for excitation manifolds is paramount to describe experiments where the pump power is high enough to enter the non-linear regime in the external field. Some common scenarios involve polariton condensation \cite{kasprzak2006bose,kena2010room,mazza2013microscopic,daskalakis2014nonlinear,plumhof2014room,grant2016efficient,pannir2022driving} and non-linear spectroscopy of molecular polaritons \cite{dunkelberger2016modified,Ebbesen2016shg,wei2018nonlinear1,Barachati2018thg,renken2021untargeted,cheng2022molecular,Cheng2022electroabsorption,hirschmann2023role}. 

As a first approximation, one may consider calculating the dynamics in higher excitation manifolds while still neglecting $1/N$ corrections of the single-molecule light-matter coupling. This regime is characterized as $N\gg N_{exc}$. For a fixed value of $N_{exc}$  and assuming a zero temperature initial state (all molecules in the electronic ground state are in the vibrational state $|\varphi^{(g)}_{1}\rangle$), the relevant Hamiltonian $\hat{H}_{0}^{(0)}(N_{exc})$ can be written as a block tri-diagonal matrix
\begin{widetext}
\begin{equation}\label{eq:hh1}
\boldsymbol{H}_{0}^{(0)}(N_{exc})=\begin{pmatrix}
\boldsymbol{H}^{(0)}_{0,N_{exc}} & \boldsymbol{V}_{0,N_{exc}} & 0 & \cdots & 0 \\
\boldsymbol{V}^{\dagger}_{0,N_{exc}} & \boldsymbol{H}^{(0)}_{0,N_{exc}-1} & \boldsymbol{V}_{0,N_{exc}-1} & \cdots & 0 \\
0 & \boldsymbol{V}_{0,N_{exc}-1}^{\dagger} & \boldsymbol{H}^{(0)}_{0,N_{exc}-2} & \cdots & 0 \\
\vdots & \vdots & \vdots & \ddots & \vdots \\
0 & 0 & 0 & \cdots & \boldsymbol{H}^{(0)}_{0,0}
\end{pmatrix}.
\end{equation} 
\end{widetext} Here, $\boldsymbol{H}^{(0)}_{0,i}$ corresponds to all states with $0$ ground state molecules vibrationally excited, and $i$ photons in the cavity mode. This is a generalization of Eq. \ref{eq:h0} for $N_{exc}>1$. Since $\boldsymbol{V}$ is the collective coupling, it cannot be treated perturbatively, leading to high computational costs for large $N_{exc}$.
%For illustration purposes, 
%\begin{widetext}
%\begin{equation}
%    \begin{pmatrix}
%N_{exc}\omega_{c} & g\langle\varphi_{1}^{(g)}|\varphi_{1}^{(e)}\rangle\sqrt{N_{exc}}\sqrt{N} & g\langle\varphi_{1}^{(g)}|\varphi_{2}^{(e)}\rangle\sqrt{N_{exc}}\sqrt{N} & \cdots & g\langle\varphi_{1}^{(g)}|\varphi_{m}^{(e)}\rangle\sqrt{N_{exc}}\sqrt{N} \\
%g\langle\varphi_{1}^{(e)}|\varphi_{1}^{(g)}\rangle\sqrt{N_{exc}}\sqrt{N} & (N_{exc}-1)\omega_{c}+\omega_{e,1} &  &  &  \\
%g\langle\varphi_{2}^{(e)}|\varphi_{1}^{(g)}\rangle\sqrt{N_{exc}}\sqrt{N} &  & (N_{exc}-1)\omega_{c}+\omega_{e,2}  &  &  \\
%\vdots &  &  & \ddots \\
%g\langle\varphi_{m}^{(e)}|\varphi_{1}^{(e)}\rangle\sqrt{N_{exc}}\sqrt{N} &  &  &  &  (N_{exc}-1)\omega_{c}+\omega_{e,m} \\
%\end{pmatrix}
%\end{equation}    
%\end{widetext}

A much more natural approach is to express the photon and matter annihilation operators as 
\begin{align}
    \hat{b}_{i}&=\sqrt{\frac{\omega_{g,i}}{2}}\left(\hat{q}_{i}+\frac{i}{\omega}\hat{p}_{i}\right),\nonumber\\
    \hat{B}_{i}&=\sqrt{\frac{\omega_{e,i}}{2}}\left(\hat{Q}_{i}+\frac{i}{\omega}\hat{P}_{i}\right),\nonumber\\
    \hat{a}&=\sqrt{\frac{\omega_{c}}{2}}\left(\hat{x}+\frac{i}{\omega}\hat{y}\right),   
\end{align} followed by the use of the Truncated-Wigner-Approximation \cite{Wigner1932on,Phuc2024semiclassical} or the Meyer-Miller mappings toolbox \cite{mmm,stmm,Ananth2007semiclassical,Vendrellmmm,Richardson2013nonadiabatic,Huo2022MMST}.

It is important to emphasize that Eq. \ref{eq:hh1} alone corresponds to a well-defined $N_{exc}$ (e.g., starting with a Fock state of photons), a situation that differs from when the initial state corresponds to a largely occupied photon coherent state, which is typically the case in experiments. Future works will focus on adapting Eq. \ref{eq:hh1} to the study of molecular polaritons in high excitation regimes that are experimentally relevant, as well as its comparison with other currently available theories that work in the same regime \cite{Piper2022PRL,Keeling2022incoherent,Pizzi2023light}.

\section{\label{sec:summ}Summary}

The collective dynamics using truncated equations (CUT-E) formalism was developed to efficiently compute the quantum dynamics of molecular polariton systems. Here, we re-derived and re-interpreted CUT-E in light of recent theoretical advances in the field. Specifically, we provided a simpler derivation using the Schwinger boson representation of the polariton Hamiltonian, which was introduced in recent studies. We also revisited the $N\rightarrow\infty$ limit, where the system can be described using classical linear optics, and for the first time showed that corrections for finite $N$ can be expressed as a $1/N$ series. This allowed us to formally establish CUT-E as a $1/N$ expansion method for the many-body dynamics of molecular polaritons. Lastly, we outlined a path forward for simulating polariton dynamics in highly excited states, enabling efficient simulations of experiments like polariton condensation and nonlinear spectroscopy.

\begin{acknowledgments}
This work was supported by the Air Force Office of Scientific Research (AFOSR) through the Multi-University Research Initiative (MURI) program no. FA9550-22-1-0317.
\end{acknowledgments}

\section*{Data Availability Statement}

Data sharing is not applicable – no new data is generated

\appendix

\section{Schwinger boson representation of molecular polaritons}

Consider a system of $N$ identical non-interacting molecules collectively coupled to a single cavity mode. The Tavis-Cummings Hamiltonian, extended to include vibrational degrees of freedom, can be written as
\begin{equation}\label{eq:hama}
\hat{H}=\sum_{i}^{N}\left(\hat{H}_{mol}^{(i)}+\hat{H}^{(i)}_{I}\right)+\hat{H}_{cav}
\end{equation}

We can represent the first-quantized many-body wavefunction of the system using a complete basis of single-particle wavefunctions $\varphi_{\nu}$ for each of the molecules in the ensemble, and the Fock state basis states $\varphi_{n_{ph}}$ for the photon mode. The time-dependent wavefunction in this basis is given by
\begin{widetext}
\begin{equation}\label{eq:vibwfa}
\Psi(q_{ph},\vec{q},t)=\sum_{n_{ph}}\sum_{\nu_{1}\nu_{2}\cdots \nu_{N}} A^{(n_{ph})}_{\nu_{1}\nu_{2}\cdots \nu_{N}}(t)\varphi_{n_{ph}}(q_{ph})\prod_{k=1}^{N}\varphi_{\nu_{k}}(q_{k}).
\end{equation}
\end{widetext}

If we impose a permutationally-symmetric initial wavefunction, such symmetry will be conserved under time evolution since the Hamiltonian is permutationally invariant. Hence, we can write
\begin{widetext}
\begin{equation}
\Psi(q_{ph},\cdots q_{\kappa}\cdots q_{\lambda} \cdots,t)=\Psi(q_{ph},\cdots q_{\lambda}\cdots q_{\kappa} \cdots,t).
\end{equation}    
\end{widetext} This implies that the dynamics of the system can be computed in a permutationally-symmetric subspace of the entire Hilbert space. An example of a permutationally-symmetric state is all molecules in the global ground state (zero temperature).
%\begin{equation}
%A^{(n_{ph})}_{\nu_{1}\cdots \nu_{\kappa}\cdots \nu_{\lambda} \cdots \nu_{N}}(t)=A^{(n_{ph})}_{\nu_{1}\cdots \nu_{\lambda}\cdots \nu_{\kappa} \cdots \nu_{N}}(t).
%\end{equation}
In terms of the bosonic symmetrization operator, we get
\begin{widetext}
\begin{equation}\label{eq:vibwf2a}
\Psi_{+}(q_{ph},\vec{q},t)=\sum_{n_{ph}}\sum_{\nu_{1}\nu_{2}\cdots \nu_{N}}\tilde{A}^{(n_{ph})}_{\nu_{1}\nu_{2}\cdots \nu_{N}}(t)\varphi_{n_{ph}}(q_{ph})\hat{S}_{+}\prod_{k=1}^{N}\varphi_{\nu_{k}}(q_{k}),
\end{equation}
\end{widetext}
where $\hat{S}_{+}$ is defined by the permanent 
\begin{equation}
    \hat{S}_{+}\prod_{k=1}^{N}\varphi_{\nu_{k}}(q_{k})=\begin{vmatrix}
\varphi_{\nu_{1}}(q_{1}) & \varphi_{\nu_{1}}(q_{2}) & \cdots  & \varphi_{\nu_{1}}(q_{N}) \\
\varphi_{\nu_{2}}(q_{1}) & \varphi_{\nu_{2}}(q_{2}) & \cdots & \varphi_{\nu_{2}}(q_{N}) \\
\vdots & \vdots & \ddots & \vdots \\
\varphi_{\nu_{N}}(q_{1}) & \varphi_{\nu_{N}}(q_{2}) & \cdots & \varphi_{\nu_{N}}(q_{N})\\
\end{vmatrix}_{+}.
\end{equation}

At this point, it is clear that we can use the second quantization formalism, just like we do for bosonic particles. This approach avoids the complex symmetrization and renormalization procedures needed in first quantization when dealing with the many-body wavefunction. Those procedures arise from asking the wrong question: what is the state of each individual molecule at any given time? Since the molecules are indistinguishable from the outset, we should instead ask how many molecules are in each single-particle state.

In the second-quantized picture this information is included in the many body states $|n_{\nu_{1}},n_{\nu_{2}},n_{\nu_{3}},\cdots\rangle$, with $n_{\nu_{i}}$ being the number of molecules occupying the state $\varphi_{\nu_{i}}$. 

To map operators to the second-quantized picture, we use bosonic creation and annihilation operators $\hat{\beta}_{\nu_{i}}$ and $\hat{\beta}^{\dagger}_{\nu_{i}}$. Analogue operators acting on the first-quantized many-body wavefunction delete or insert a single-particle state $\varphi_{\nu_{i}}$ while conserving its symmetrization. Acting on the second-quantized many-body wavefunction, $\hat{\beta}_{\nu_{i}}$ and $\hat{\beta}^{\dagger}_{\nu_{i}}$ will lower and increase the occupation in the single-particle state $\varphi_{\nu_{i}}$ by one, respectively; they destroy and create \textit{molecules} in the single-particle states $\varphi_{\nu_{i}}$. Moreover, we can map any symmetric sum of single-particle operators, $\hat{O}=\sum_{k=1}^{N}\hat{o}^{(k)}$, from the first-quantized to the second-quantized picture, using the creation and annihilation operators via \cite{Bruus2004many}
\begin{equation}\label{eq:map}
\hat{O}\rightarrow\sum_{\nu_{i}\nu_{j}}\langle\nu_{i}|\hat{o}|\nu_{j}\rangle\hat{\beta}^{\dagger}_{\nu_{i}}\hat{\beta}_{\nu_{j}}.
\end{equation}
Using the vibronic eigenstates $|\varphi^{(g)}_{i},g\rangle$ and $|\varphi^{(e)}_{i},e\rangle$ as the basis set, and the mapping between first- and second-quantized operators above, we can obtain the molecular polariton Hamiltonian in Eq. \ref{eq:ham2} of the manuscript.

\section{Block-tridigonal structure of the molecular polariton Hamiltonian}

The molecular polariton Hamiltonian in Eq. \ref{eq:ham2} can be partitioned as 
\begin{align}
&\hat{H}=\hat{H}^{(0)}+\hat{v},\nonumber\\
&\hat{H}^{(0)}=\omega_{c}\hat{a}^{\dagger}\hat{a}+\sum_{i}^{m}\omega_{g,i}\hat{b}_{i}^{\dagger}\hat{b}_{i}+\sum_{i}^{m}\omega_{e,i}\hat{B}_{i}^{\dagger}\hat{B}_{i}\nonumber\\
&+g\sum_{i}^{m}\left(\langle\varphi^{(e)}_{i}|\varphi^{(g)}_{1}\rangle\hat{B}^{\dagger}_{i}\hat{b}_{1}\hat{a}+\langle\varphi^{(g)}_{1}|\varphi^{(e)}_{i}\rangle\hat{B}_{i}\hat{b}^{\dagger}_{1}\hat{a}^{\dagger}\right),\nonumber\\
&\hat{v}=g\sum_{i,j>1}^{m}\left(\langle\varphi^{(e)}_{i}|\varphi^{(g)}_{j}\rangle\hat{B}^{\dagger}_{i}\hat{b}_{j}\hat{a}+\langle\varphi^{(g)}_{j}|\varphi^{(e)}_{i}\rangle\hat{B}_{i}\hat{b}^{\dagger}_{j}\hat{a}^{\dagger}\right),
\end{align} where $\hat{H}^{(0)}$ contains the collective light-matter couplings while the perturbation $\hat{v}$ accounts for all single-molecule light-matter couplings. 

Notice that $\hat{H}^{(0)}$ commutes with the operators $\hat{n}_{k>1}=\hat{b}^{\dagger}_{k>1}\hat{b}_{k>1}$, i.e., the number of molecules in the electronic ground state in each vibrational excited state. This implies that $\hat{H}^{(0)}$ can be written as a block-diagonal matrix of the form
\begin{equation}    \hat{H}^{(0)}=\bigoplus_{n_{k}=0}^{N}\hat{H}^{(0)}_{n_{k}}.
\end{equation} Likewise, $\hat{H}^{(0)}_{n_{k}}$ can also be written as a block-diagonal matrix where each block represents a distribution of the $n_{k}$ molecules amongst all excited $m-1$ vibrational excited states. For example, for $n_{k}=1$ we have
\begin{equation}
    \hat{H}^{(0)}_{1}=\bigoplus_{k>1}^{m}\hat{H}^{(0)}_{1_{k}}.
\end{equation}

Notice that $\hat{n}_{k>1}$ does not commute with the perturbation $\hat{v}$. Based on this, we refer to the number of ground state molecules in each vibrationally excited state as a \textit{quasi-conserved quantity}. In other words, collective light-matter coupling cannot create or destroy phonons in electronic ground-state molecules. Finally, notice that the perturbation $\hat{v}$ can only create or destroy one ground-state molecule with vibrational excitations at a time. This gives the molecular polariton Hamiltonian in Eq. \ref{eq:blocks} its block-tridiagonal structure.

\bibliography{main}% Produces the bibliography via BibTeX.

\end{document}